\documentclass[a4paper]{article}

\usepackage{ISCSLP2022}
\usepackage{color}
\usepackage{hyperref}
\usepackage{stfloats}

\title{Robust MelGAN: A robust universal neural vocoder for high-fidelity TTS}
\name{Kun Song$^1$, Jian Cong$^1$, Xinsheng Wang$^1$, Yongmao Zhang$^1$, Lei Xie$^{1,*}$, Ning Jiang$^2$, Haiying Wu$^2$}
\address{
  $^1$Audio, Speech and Language Processing Group (ASLP@NPU), School of Computer Science, \\Northwestern Polytechnical University, Xi'an, China\\ $^2$Mashang Consumer Finance Co., Ltd.}

\email{kunsong.npu.se@gmail.com, lxie@nwpu.edu.cn, ning.jiang02@msxf.com}

\begin{document}

\maketitle

\begin{abstract}
 In current two-stage neural text-to-speech (TTS) paradigm, it is ideal to have a \textit{universal} neural vocoder, once trained, which is robust to imperfect mel-spectrogram predicted from the acoustic model. To this end, we propose Robust MelGAN vocoder by solving the original multi-band MelGAN's metallic sound problem and increasing its generalization ability. Specifically, we introduce a fine-grained network dropout strategy to the generator. With a specifically designed over-smooth handler which separates speech signal intro periodic and aperiodic components, we only perform network dropout to the aperodic components, which alleviates metallic sounding and maintains good speaker similarity. To further improve generalization ability, we introduce several data augmentation methods to augment fake data in the discriminator, including harmonic shift, harmonic noise and phase noise. Experiments show that Robust MelGAN can be used as a universal vocoder, significantly improving sound quality in TTS systems built on various types of data. \footnote{Audio samples are available at \url{https://RobustMelGAN.github.io/RobustMelGAN/}}
\end{abstract}
\renewcommand{\thefootnote}{\fnsymbol{footnote}}
\footnotetext{Lei Xie is the Corresponding author.}

\vspace{-5pt}
\section{Introduction}
A modern text-to-speech (TTS) system is composed of a neural acoustic model (AM) and a neural vocoder, where mel-spectrogram and its variants typically serve as the intermediate representation. Specifically, the neural vocoder takes the AM predicted mel-spectrogram as input and generates the waveform. To ensure better quality for service, both models are trained with sizable studio-recorded high-quality data from a target speaker, and sometimes ground-truth aligned (GTA) mel-spectrogram is adopted to fine-tune the neural vocoder to alleviate the mismatch between ground truth and predicted mel-spectrogram. A \textit{universal vocoder} aims to abandon such a time- and resource-consuming procedure and produce the high-quality waveform from the intermediate representation (e.g. mel-spectrogram) that may come from an arbitrary speaker hopefully with an arbitrary speaking style. In other words, the neural vocoder, once trained with data from a multiple speaker inventory, is robust enough to produce high-quality speech from the spectrogram that belongs to a new (\textit{unseen}) speaker.

Although traditional digital signal processing (DSP) based source-filter vocoders~\cite{DBLP:journals/speech/KawaharaMC99,DBLP:journals/ieicet/MoriseYO16} are \textit{universal} in that sense without model training, the quality of the produced audio is poor in general with mechanical sounding. With the help of deep learning-based data-driven approach, a \textit{neural vocoder} can produce high-fidelity sounding with even human parity quality.

The mainstream neural vocoders can be categorized into autoregressive (AR)~\cite{DBLP:conf/ssw/OordDZSVGKSK16, DBLP:conf/icml/KalchbrennerESN18, DBLP:conf/icassp/ValinS19} and non-AR models. Non-AR models have become more popular as they have a faster inference speed with parallel generation ability in nature.
 In the non-AR category, flow-based models~\cite{DBLP:conf/icassp/PrengerVC19, DBLP:conf/icml/PingPZ020} require a longer training time, and the diffusion probabilistic models~\cite{DBLP:conf/iclr/KongPHZC21} have a relatively slow inference speed. In contrast, vocoders based on generative adversarial network (GAN)~\cite{DBLP:conf/nips/KumarKBGTSBBC19, DBLP:conf/icassp/YamamotoSK20, DBLP:conf/nips/KongKB20} have apparent advantages -- it can produce high-quality waveform with a high inference speed. GAN-based vocoders rely on adversarial learning of two sub-networks: a generator, which generates samples that try to fool the discriminator; a discriminator, which tries to distinguish the difference between the fake sample generated by the generator and the real sample. The generator implicitly reconstructs the phase information, generates speech waveform through spectrogram in parallel, and optimizes the modeling process through the discriminator.

In this paper, we aim to design a robust GAN-based universal neural vocoder for high-fidelity TTS. This task is not trivial become of the following two aspects.
First, the problem lies in the nature and design of current GAN-based vocoders. Unlike the AR models, a GAN vocoder does not use the previously generated sample as prior signal information, nor does it explicitly distinguish periodic and aperiodic components of signals like the source-filter models. When encountering the over-smoothed spectrogram generated by an acoustic model, the upsampling network may regard the aperiodic signal components in the spectrogram as periodic signal components, thus generating harmonic-like artifacts that like metallic sounding. Second, the generalization ability of GAN vocoders is poor. That is to say, they have insufficient ability to generate quality waveform using the spectrogram from unseen speakers as well as low-quality acoustic models, which leads to substantial degradation of sound quality. Some previous works have attempted to solve these problems. For the first problem targeting to the over-smoothed spectrogram, improving the generalization ability of the acoustic model is clearly beneficial~\cite{DBLP:conf/acl/0006TQZL22}. For the generalization ability of the vocoder itself, adopting a discriminator in the frequency domain is proven to be useful for improving sound quality of high-frequency components of speech and unseen speakers~\cite{DBLP:conf/interspeech/JangLYKK21}.

In this paper, we propose \textit{Robust MelGAN} as a universal neural vocoder for high-fidelity TTS. Specifically, we build on multi-band MelGAN~\cite{DBLP:conf/slt/YangYLF0X21}, a fast and lightweight GAN vocoder, and address the above two problems by 1) adopting a fine-grained network dropout strategy in the generator, which improves robustness on the prediction of the aperiodic signals; and 2) data augmentation tricks to the discriminator, including harmonic shift, harmonic noise, and phase noise, adopted to generate fake samples. Experiments show that Robust MelGAN can significantly improve sound quality in TTS systems built on various types of data, including studio-recorded multi-speaker and expressive speech as well as low-quality speech in few-shot speaker adaption tasks.


\vspace{-5pt}
\section{Method}
Updated from MelGAN~\cite{DBLP:conf/nips/KumarKBGTSBBC19}, multi-band MelGAN~\cite{DBLP:conf/slt/YangYLF0X21} is a popular lightweight GAN vocoder using sub-band modeling. It takes mel-spectrogram as input to generate signals in multiple frequency bands instead of the full frequency band in MelGAN, and the sub-band signals are summed back to full-band audio signal through Pseudo Quadrature Mirror Filterbank (PQMF). For the generator, a stack of transposed convolution is adopted to upsample the mel-spectrogram to match the frequency of waveforms. Each transposed convolution is followed by a stack of residual blocks. Compared with~\cite{DBLP:conf/icassp/YamamotoSK20} and ~\cite{DBLP:conf/nips/KongKB20}, multi-band MelGAN's serial residual blocks enable it to maintain relatively good sound quality and better inference speed. For the discriminator, it uses full-band signals as input and multiple discriminators to discriminate features of different scales. 

Despite the advantages, similar to other GAN-vocoders, multi-band MelGAN still suffers from the metallic sounding problem mainly caused by inappropriate processing of aperiodic signals as well as the robustness problem which means it does not generalize well to unseen or degraded mel-spectrogram. In this paper, we aim to update multi-band MelGAN to Robust MelGAN which is more appropriate to be used as a universal vocoder.
\vspace{-5pt}
\subsection{Generator with dropout}
\begin{figure}[ht]
\setlength{\abovecaptionskip}{0.1cm}
	\centering
	\includegraphics[width=1.02\linewidth]{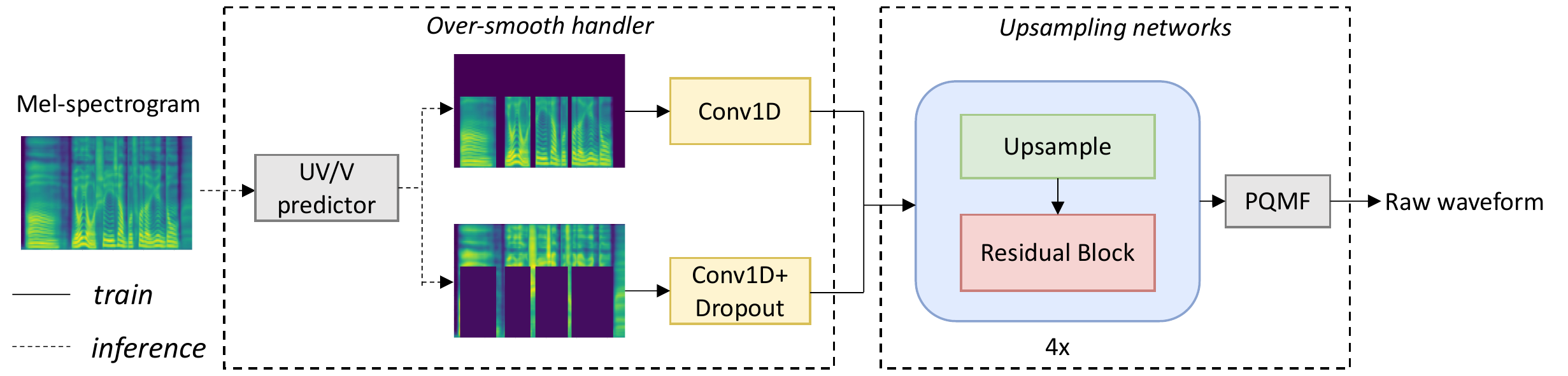}
	\caption{
		Architecture of generator with over-smooth handler.
	}
	\label{model_flatts}\vspace{-4pt}
\end{figure}

\textbf{Metallic Sounding Problem} Metallic sounding is characterized by transverse harmonic artifacts, which sound like metal friction and seriously affect sound quality. This phenomenon is mainly caused by the \textit{over-smooth} of the spectrogram generated by the acoustic model, where the signal's aperiodic components will show characteristics similar to periodic signals.  However, since the MelGAN generator uses upsampling networks to model the phase without prior knowledge of the signal, it will generate a pseudo-harmonic signal that looks like harmonics in spectrum and sounds like metallic friction. Metallic sounding is particularly evident in unvoiced segment and high-frequency components of expressive speech because of the rich aperiodic signals caused by high-intensity oral refraction. Therefore, the way to alleviate metallic sounding is to improve the robustness on the prediction of aperiodic signals. As shown in Figure~\ref{model_flatts}, we follow the generator architecture of multi-band MelGAN and particularly propose a \textit{over-smooth handler} to avoid metallic sounding in the unvoiced and high-frequency parts of speech.

\vspace{-5pt}
\textbf{Over-smooth Handler} As we know, \textit{dropout}~\cite{DBLP:journals/jmlr/SrivastavaHKSS14} is widely used in neural networks to prevent over-fitting and improve generalization ability. We can also use dropout in prenet of the generator to alleviate the above-mentioned over-smooth problem of predicted mel-spectrogram and improve generalization ability to unseen speakers as well. Specifically, by randomly masking some neurons in prenet, we can make the waveform generated by the over-smoothed mel-spectrogram close to the waveform generated by the real mel-spectrogram. However, we only want to apply dropout to aperiodic components of the mel-spectrogram because we find that dropout of the periodic components affects the fundamental and formant frequency of speech, hindering speaker similarity. This problem can be alleviated by empirically tuning the dropout rate, e.g., adopting a small dropout rate. But specific configuration depends on data and a small rate leads to insignificant improvement. 

To fully solve the problem, we roughly separate speech into periodic and aperiodic components, and use two separate prenets and perform dropout only to the specific prenet handling the aperiodic components. To be specific, for the unvoiced segments, since there is no periodic components, we can adopt network dropout to the full-band signal; for the voiced segments, we separate mel-spectrogram into low- and high-frequency components, and only adopt network dropout to high-frequency components where aperiodic components dominate. To this end, as shown in Figure~\ref{model_flatts}, we introduce an \textit{over-smooth handler} as a pre-processing step before the upsampling networks. Specifically, in the over-smooth handler, we first separate mel-spectrogram of the voiced segment into low- and high-frequency components. Then we use mel-spectrogram as input to predict UV/V by a UV/V predictor and separate unvoiced and voiced segments by the UV/V mask. The UV/V predictor consists of several Conv1D layers and it is trained separately. After then, we adopt dropout to the prenet of the high-frequency components and unvoiced segments while going through the Conv1D layers of the prenet.

\vspace{-5pt}
\subsection{Discriminator with data augmentation}
Data augmentation has been previously adopted in speech generation tasks to improve generalization ability. In this paper, we design several data augmentation strategies used in the discriminator. Specifically, we add perturbations to raw data $x$ to generate augmented data $x'$ as fake data and let the model to distinguish real/fake from $x$ and $x'$. 


\begin{figure}[ht]
\setlength{\abovecaptionskip}{0.1cm}
	\centering
	\includegraphics[width=7.5cm,height=5.5cm]{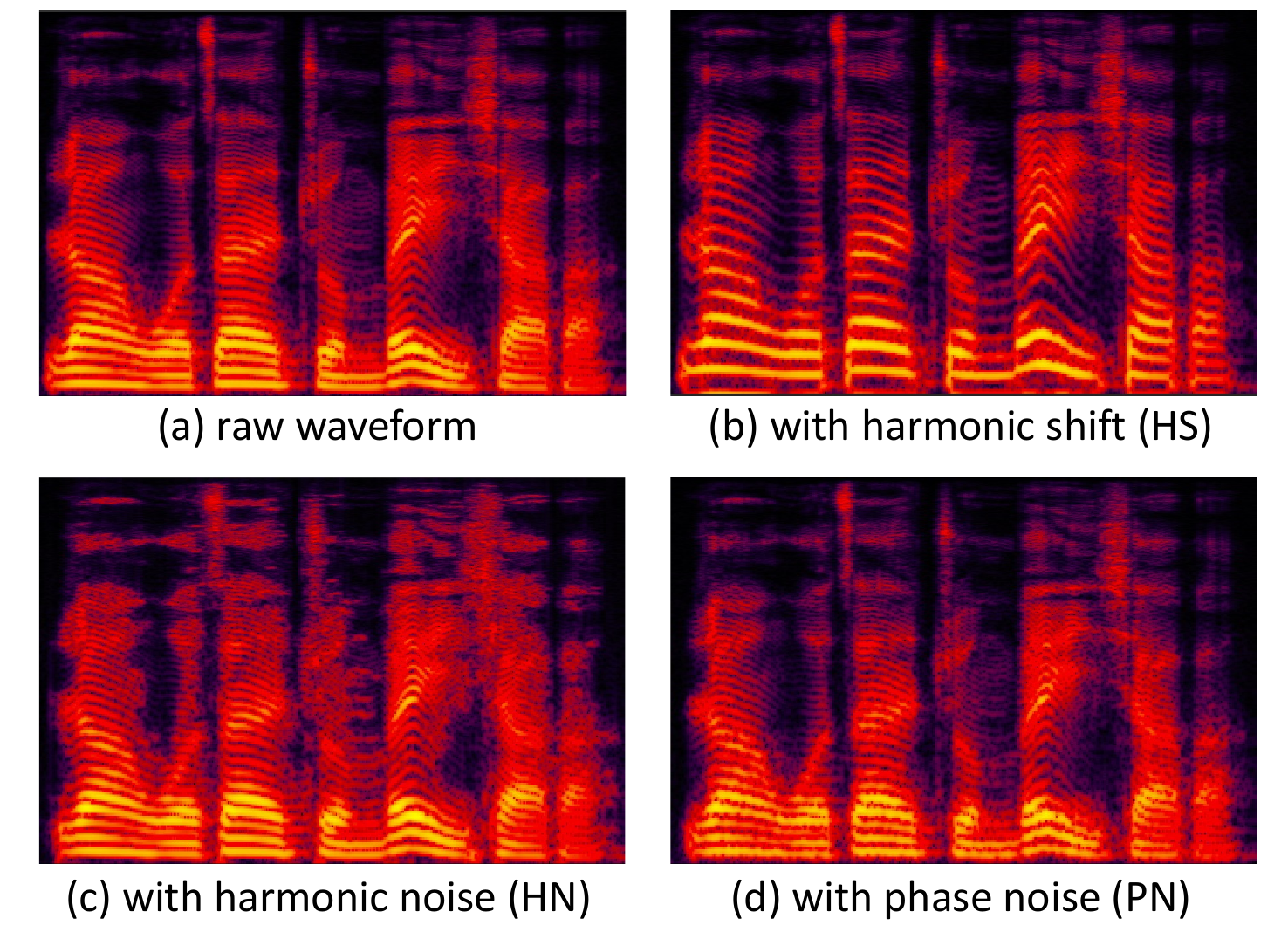}
	\caption{
		The spectrogram of raw and augmented data.
	}
	\label{augment}
\end{figure}
\vspace{-8pt}
\textbf{Harmonic shift (HS)} Harmonics are the main components of speech, composed of fundamental and formant frequency, which determines mostly the timbre of speech and its quality. In order to improve the model's generalization ability, we simulate unseen speakers' data through timbre disturbance. As shown in Figure~\ref{augment}(b), we modify the $F0$ of the raw data and scale the formant to change its timbre to obtain new data. To be specific, we use the parselmouth~\cite{DBLP:journals/jphonetics/JadoulTB18} to implement the harmonic shift and randomly select formant shift ratio, median, and pitch range factor in the range of 0.9\textasciitilde1.1, 100\textasciitilde500 and 0.8\textasciitilde1.2, respectively.

\textbf{Harmonic noise (HN)} Speech is composed of both periodic and aperiodic components.  The aperiodic component may help in characterizing perceived attributes of voice quality including breathness or roughness. While periodic signal dominates the voiced part of speech, some random component is also present in the voiced part, including those wrapped around the harmonics. Besides voiced fricatives and breathy vowels, random component can appears in normal vowels because of turbulence
of air around the instant of glottal closure, resulting in aspiration noise. Previous studies show that appropriate addition of the aperiodic component in voiced excitation is essential to natural synthetic speech.

When the prediction accuracy of the aperiodic component is low, sound quality will be degraded. To solve this problem, we add some extra random noise around the harmonics to simulate the fake examples. Figure~\ref{augment}(c) shows an example. The procedure of adding harmonic noise is as follows.


\begin{equation}
\begin{aligned}
& F0, sp,ap =\text{Analysis}(x); \\
&s p'=\left
\{\begin{array}{l}
s p, s p<\alpha  \\
s p+\beta * noise, s p \geqslant \alpha ;
\end{array}\right. \\
&x'=\text{Synthesis}\left(F 0, s p', a p\right) ;
\end{aligned}
\end{equation}
where $x$ represents the original audio. We use WORLD~\cite{DBLP:journals/ieicet/MoriseYO16} to extract fundamental frequency $F0$,  periodic signal $sp$, and aperiodic signal $ap$ from $x$. Since the part of $sp$ with a higher value can represent the harmonics of speech, we increase the $noise$ of U(0, 1) with weight $\beta$ from the part greater than $\alpha$ to obtain $sp'$. We randomly select $\alpha$ and $\beta$ in the range of [1e-4, 5e-4, 1e-3] and [1e-5, 3e-5, 5e-5, 8e-5], respectively. Finally we use WORLD to re-synthesise the disturbed waveform $x'$ from $F0$, $sp'$ and $ap$.

\textbf{Phase noise (PN)} Vocoders based on upsampling networks implicitly model the phase of speech signal.
Phase mismatch is a major problem between synthetic speech and real speech, leading to lower quality in the synthetic speech.
We simulate this kind of mismatch and treat the the augmented data with phase noise as fake samples. The process is described as follows. Figure~\ref{augment}(d) shows an example with added phase noise.

\begin{equation}
\begin{aligned}
& magnitude, phase  = \text{STFT} (x) ; \\
&phase' = phase +\alpha * noise; \\
&x'= \text{iSTFT}(magnitude, phase'); 
\end{aligned}
\end{equation}
where $x$ represents the original audio. We use short-time Fourier transform (STFT) to extract the $magnitude$ and $phase$ of $x$, and then $noise$ of U(0, 1) with weight of $\alpha$ is added to $phase$ to obtain $phase'$. We randomly select $\alpha$ in the range of 0.5\textasciitilde1.5, with a common difference of 0.1. At last, we use iSTFT to generate the disturbed waveform $x'$ by the $magnitude$ and $phase'$.
\vspace{-5pt}
\subsection{Training loss}
Our model has the same training process as multi-band MelGAN~\cite{DBLP:conf/slt/YangYLF0X21}. But differently, since we adopt augmented data, we need to make extra real/fake discrimination. Hence our discriminator loss is
\vspace{-5pt}
\begin{equation}
\mathcal{L}_{D}=\mathbb{E}_{x}\left[\left(D(x)-1\right)^{2}\right] + \mathbb{E}_{\hat{x}}\left[D(\hat{x})^{2}\right]
 + \mathbb{E}_{x'}\left[D(x')^{2}\right]
\end{equation}
where $x$, $\hat{x}$ and $x'$ represent the ground-truth data, data generated by the generator, and the augmented data.






\section{Experiments}

We first separately evaluate the proposed methods in the generator and the discriminator, and then perform overall evaluation and ablation study. We use PESQ~\cite{DBLP:conf/icassp/RixBHH01} and $F0$ Root Mean Square Error ($F0$-RMSE) for objective evaluation. For subjective evaluation, we adopt mean opinion score (MOS) tests to investigate the performance of the proposed methods. There are 20 Mandarin listeners evaluating the speech quality.

\vspace{-8pt}
\subsection{Datasets}
For the training of the universal vocoder, we use an internal studio quality Mandarin dataset which consists of 309 speakers with 622,679 utterances in total. We conduct both copysyn and TTS tests. For the vocoder \textit{copysyn} test, we reserve 50 utterances of 10 speakers randomly selected from the training data as the \textit{seen} speakers. We use another multi-speaker Mandarin dataset AISHELL-3~\cite{DBLP:conf/interspeech/ShiBXZL21} as the \textit{unseen} speakers, from which we randomly selected 50 utterances of 10 speakers for copysyn test. Note that AISHELL-3 is composed of recorded speech from ordinary speakers in a typical room with inevitable ambient noise and small reverberation. 

For TTS evaluation, we first train a multi-speaker DelightfulTTS~\cite{DBLP:journals/corr/abs-2110-12612} using the same set of data as the vocoder training. Then we finetune this model using different sets of data listed below for TTS tests. Note that the same universal vocoder is adopted no matter which acoustic model is used.
\begin{itemize}
    \item \textbf{Hifi}: we use a high-fidelity dataset consisting of a male speaker and a female speaker with typical reading style, each with 5,000 utterances. This test represents the high resource scenario where sizable high-fidelity data of target speaker is available.
    \item \textbf{Adapt}: we use 3 male and 2 female speakers, each with 50 utterances recorded in a typical office room, to finetune the base model for low resource speaker adaption.
    \item \textbf{Spon:} we use a studio-recorded conversation dataset which comprises a male speaker and a female speaker, each with 8,000 utterances. The dataset contains expressive conversational speech with fast talking, different emotions, breaths and even smiles. 
\end{itemize}

 
 All samples are resampled to 24Khz, and we extract 80-bands log-mel-spectrogram with a 1024-point FFT, 256 sample frame shift and 1024 sample frame length. We use 50 sentences, including short and long sentences, for TTS tests.

\vspace{-8pt}
\subsection{Model Configuration}

We use multi-band MelGAN~\cite{DBLP:conf/slt/YangYLF0X21}as our baseline, which has 3M model parameters with a real-time factor (RTF) of 0.06 on an Intel Xeon E5 CPU. For the generator, upsample factors and channels are set to [2, 2, 4, 4] and [384, 192, 128, 64, 32] respectively, prenet uses three layers of Conv1D, and other configurations follows those in multi-band MelGAN. We use a multi-resolution spectrogram discriminator in all experiments for the discriminator, and the configurations are consistent with those in UnivNet~\cite{DBLP:conf/interspeech/JangLYKK21}. For our proposed Robust MelGAN, UV/V predictor has 4 layers of Conv1D and hidden set to[256, 256, 256, 256]. We use the $F0$ value extracted from Harvest~\cite{DBLP:conf/interspeech/Morise17} with Stonemask to determine UV/V, where the frame with value 0 in $F0$ is regarded as unvoiced segment and the rest as voiced segment. We set the lower 50 dimensions of mel-spectrogram as low-frequency and the rest 30 as high-frequency components, and dropout is adopted in the second layer of the prenet. In all the experiments, dropout rate is set to 0.5. With the addition of the UV/V predictor and an extra prenet over the oracle multi-band MelGAN, Robust MelGAN has 4.7M model parameters but RTF remains at 0.06 on the same GPU as the extra computation induced by the UV/V predictor and prenet is negligible. All the vocoders are trained to 1.5M steps with batch size 32. For TTS, we use DelightfulTTS as our acoustic model and follow the configuration in its original paper~\cite{DBLP:journals/corr/abs-2110-12612}.

%
%
%

\begin{table}[]
    
	\centering
	\caption{The MOS test of the proposed methods to the generator. Here + means with and - means without.}
	\setlength{\tabcolsep}{3mm}{
	\begin{tabular}{lcl}
		\toprule
		Model &MOS  \\ 
		\midrule
		
		Multi-band MelGAN &3.19$\pm$0.11  \\
        \ + over-smooth handler   &\textbf{3.78$\pm$0.10}  \\
        \ \ \quad- unvoiced segment dropout &3.55$\pm$0.10 \\
        \ \ \quad- voiced segment dropout &3.60$\pm$0.08 \\
        \ + full-band dropout   &3.01$\pm$0.07  \\
        
		\bottomrule
	\end{tabular}
	}
	\label{MOS for istft}\vspace{-5pt}
\end{table}

\vspace{-8pt}

\subsection{Experiment on generator}
\begin{figure}[ht]
\setlength{\abovecaptionskip}{0.1cm}
	\centering
	\includegraphics[width=7.26cm,height=2.64cm]{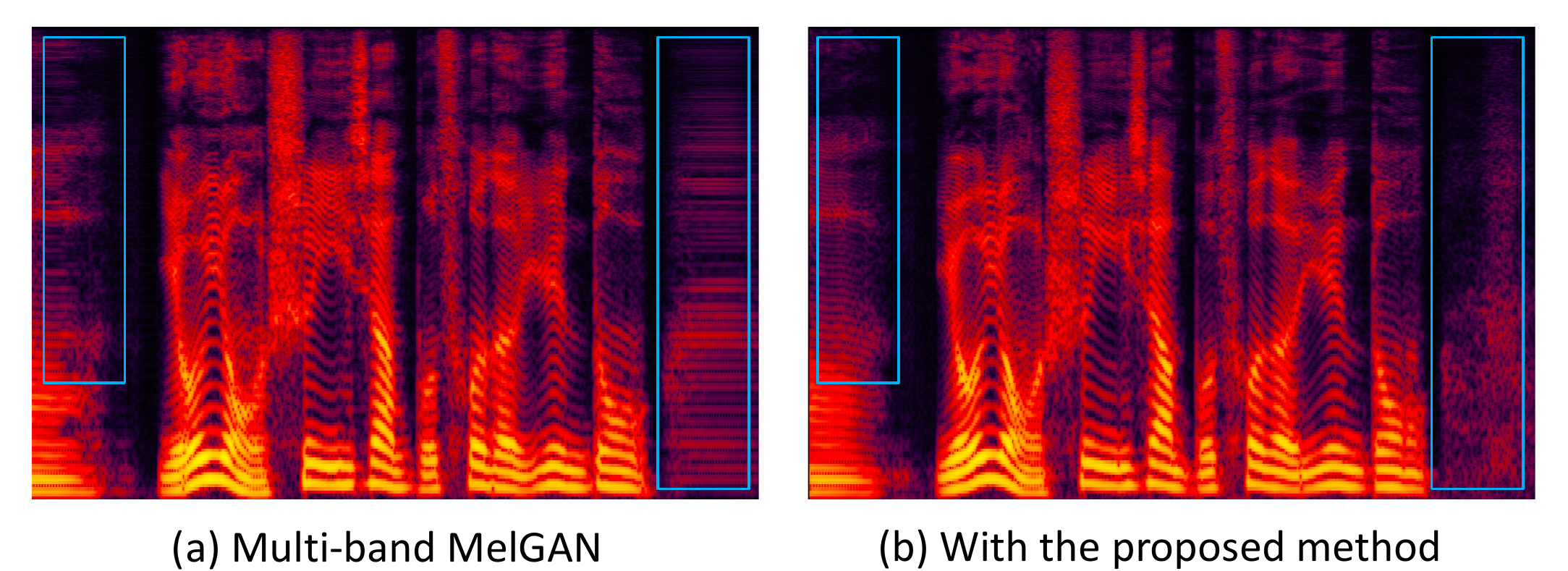}
	\caption{
		Mel-spectrogram of a synthetic sample before (a) and after (b) using the proposed method, where the metallic sounding problem in the boxed areas is solved.
	}
	\label{istftmodel}\vspace{-10pt}
\end{figure}

\vspace{-2pt}
Since the proposed method in the generator is mainly designed to solve metallic sounding caused by the over-smooth problem of the acoustic model, we particularly conduct TTS experiment on the expressive \textit{Spon} data. MOS results are shown in Table~\ref{MOS for istft}.  As can be seen, the use of the proposed over-smooth handler leads to significant MOS improvement. An example of synthetic speech is shown Figure~\ref{istftmodel}. We can see from the mel-spectrogram, the problem in the blue boxes in (a), which induces unpleasant metallic sounding, has been solved in (b). The aperiodic components in the blue boxes are high-frequency signal and breath. We also notice that applying network dropout to voiced and unvoiced segments only also beneficial. In contrast, network dropout to the entire full-band signal is clearly harmful. Listeners reflect that speaker similarity will be seriously affected.

\begin{table}[]
    
	\centering
	\caption{The MOS test on the use of data augmentation in the discriminator.}\vspace{-4pt}
	\setlength{\tabcolsep}{3mm}{
	\begin{tabular}{lcl}
		\toprule
		Model &MOS  \\ 
		\midrule
		
		Multi-band MelGAN &3.49$\pm$0.11  \\
        \ +HS   &3.65$\pm$0.08  \\
        \ +HN   &3.56$\pm$0.10  \\
        \ +PN   &3.61$\pm$0.09  \\
        \ +HS/HN/PN (Random pick)   &3.63$\pm$0.11  \\
        \ +HS+HN+PN (Use all)   &\textbf{3.72$\pm$0.12}  \\
		\bottomrule
	\end{tabular}
	}
	\label{MOS for data}\vspace{-10pt}
\end{table}

\begin{table*}[bp]

\centering
	\caption{Evaluation on multi-band MelGAN, Robust MelGAN and Ablations. `-' means without.}
\resizebox{\linewidth}{!}{ 
\begin{tabular}{l|ccc|ccc|c|c|c}
\toprule
      & \multicolumn{3}{c|}{Seen speaker (copysyn)} & \multicolumn{3}{c|}{Unseen speaker (copysyn)} & Hifi (TTS) & Adapt (TTS) & Spon (TTS) \\ \midrule
Model & MOS    & PESQ    & $F0$-RMSE   & MOS     & PESQ    & $F0$-RMSE    & MOS  & MOS   & MOS  \\ \midrule
Multi-band MelGAN     &    3.93$\pm$0.11    &  3.21       &  10.71      &     3.55$\pm$0.11    &   2.89      &    19.54     &    3.69$\pm$0.09  &3.54$\pm$0.11       &  3.28$\pm$0.10    \\
Robust MelGAN     &     3.94$\pm$0.11   &    3.23     &  \textbf{7.81}      &     3.63$\pm$0.12    &  \textbf{2.94}       &     \textbf{18.25}    &   \textbf{3.82$\pm$0.11}   &  \textbf{3.76$\pm$0.10}     &   \textbf{3.75$\pm$0.12}   \\ 
\ - over-smooth handler    & \textbf{3.97$\pm$0.10}       &    \textbf{3.28}     &  12.26      &     \textbf{3.65$\pm$0.11}    &    2.94     &   18.26      &   3.81$\pm$0.13   &     3.74$\pm$0.12  & 3.49$\pm$0.11     \\
\ - data augmentation     &   3.89$\pm$0.12     &   3.18  &     7.99   &    3.43$\pm$0.13     &  2.90       &   18.34      &  3.68$\pm$0.09    &   3.52$\pm$0.11    & 3.61$\pm$0.11     \\\midrule
Recording     &   4.42$\pm$0.09     &    4.50     &  0.00      &   4.31$\pm$0.10      &   4.50      &   0.00      &  -    &   -    &-      \\ \bottomrule

\end{tabular}
}
\label{MOS test}
\end{table*}

\vspace{-10pt}
\subsection{Experiment on discriminator}
\vspace{-2pt}
We test our data augmentation strategies applied to the discriminator using \textit{Adapt} dataset. Here low resource speaker adaptation task is particularly used to evaluate robustness of our universal vocoder.
 Experimental results are shown in Table~\ref{MOS for data}. It can be seen that all the three methods are useful to sound quality, with harmonic shift playing the most significant role. In addition, adopting all the augmentation methods is more effective than using the three randomly.
\vspace{-8pt}

\subsection{Evaluation on entire Robust MelGAN}
\vspace{-2pt}
TTS experiments in Section 3.3 and 3.4 have shown the effectiveness of the proposed improvements on the generator and discriminator. Here we further evaluate the entire Robust MelGAN with more comprehensive experiments and ablation study. For copysyn, we conduct both objective and subjective experiments on seen and unseen speakers. For TTS, we perform subjective evaluation on three different sets of data -- Hifi, Adapt and Spon. Results in Table~\ref{MOS test} show that Robust MelGAN achieves clearly better MOS scores in different TTS tasks as compared with the oracle Multi-band MelGAN. The improvement is more evident with the increase of the task difficulty ($\text{Hifi}\rightarrow\text{Adapt}\rightarrow\text{Spon}$). For the copysyn task, while the entire Robust MelGAN is slightly inferior to that without the over-smooth handler, it still outperforms the original Multi-band MelGAN, especially for the unseen speakers. The reason behind is that the over-smooth handler is mainly introduced to handle the over-smooth problem that leads to poor quality of aperiodic signals in synthetic speech while this problem does not exist in the ground-truth spectrogram and the use of dropout will induce some noise in the re-synthesized speech of the original speech. In summary, Robust MelGAN shows it robustness as a universal vocoder to the TTS systems trained using various types of data, with significant quality improvement as compared with the original multi-band MelGAN.

%
%


\vspace{-10pt}
\section{Conclusions}
In this paper, we propose Robust MelGAN, aiming to make GAN-based vocoder to meet the requirements of a universal vocoder. To solve the problem of metallic sounding, we adopt a fine-grained network dropout strategy in the generator. In addition, several data augmentation methods are introduced to the discriminator to improve generalization ability of the vocoder. Experimental results show that the proposed Robust MelGAN shows superior performance in various TTS tasks with better sound quality and robustness.

\bibliographystyle{IEEEtran}

\bibliography{mybib}


\end{document}